\documentclass[a4paper,12pt]{article}

\usepackage{epsfig,graphicx,amsmath,amssymb}
\usepackage{url}
\usepackage{hyperref}
\newcommand{\ee}{e^{+} e^{-}}
\def\amu{a_\mu}

\def \zc {Z_c}

\def \ee {e^+e^-}
\def \jp {J/\psi}
\def \ppj {\pi^+\pi^- J/\psi}
\newcommand{\jpsi}{J/\psi}
\newcommand{\elp}{e^{+}}
\newcommand{\elm}{e^{-}}
\newcommand{\mup}{\mu^{+}}
\newcommand{\mum}{\mu^{-}}
\newcommand{\pip}{\pi^{+}}
\newcommand{\pim}{\pi^{-}}
\newcommand{\piz}{\pi^{0}}

\begin{document}

\thispagestyle{empty}

$\phantom{.}$

\hfill

\begin{center}
{\Large {\bf Constraining the Hadronic Contributions to the Muon Anomalous Magnetic Moment} \\
\vspace{0.75cm}}

\vspace{1cm}

{\large April 10--12, 2013 in Trento, Italy}

\vspace{2cm}

{\it Editors:}
P.~Masjuan (Mainz) and G.~Venanzoni (Frascati)

\vspace{2.5cm}

ABSTRACT

\end{center}

\vspace{0.3cm}

\noindent
The mini-proceedings of the Workshop on "Constraining the hadronic contributions to the muon anomalous magnetic moment" which included the "13th meeting of the Radio MonteCarLow WG" and the "Satellite meeting R-Measurements at BES-III" held in Trento from April 10$^{\rm th}$ to 12$^{\rm th}$, 2013, are presented. This collaboration meeting aims to bring together the experimental $e^+e^-$ collider communities from BaBar, Belle, BESIII, CMD2, KLOE, and SND, with theorists working in the fields of meson transitions form factors, hadronic contributions to $(g-2)_{\mu}$ and effective fine structure constant, and development of Monte Carlo generator and Radiative Corrections for precision $\ee$ and $\tau$ physics.

\medskip\noindent
The web page of the conference, which contains all talks, can be found at
\begin{center}
\url{https://agenda.infn.it/conferenceDisplay.py?confId=6104}
\end{center}

\vspace{0.5cm}

\noindent
We acknowledge the support of Deutsche Forschungsgemeinschaft DFG through the Collaborative Research Center ``The Low-Energy Frontier of the Standard Model" (SFB 1044), and the ECT* for the warm hospitality and support to the meeting. 

\noindent
This work is a part of the activity of the "Working Group on Radiative Corrections and Monte Carlo Generators for Low Energies" [\url{http://www.lnf.infn.it/wg/sighad/}].

\newpage

{$\phantom{=}$}

\vspace{0.5cm}

\tableofcontents

\newpage

\section{Introduction to the Workshop}

\addtocontents{toc}{\hspace{1cm}{\sl G.~Venanzoni {\it et al.}}\par}

\vspace{5mm}

\noindent
H.~Czyz$^1$, A.~Denig$^2$, M.~Vanderhaeghen$^2$, and G.~Venanzoni$^3$

\vspace{5mm}

\noindent
$^1$ University of Silesia, Katowice, Poland\\
$^2$ Institut f\"ur Kernphysik, Johannes Gutenberg-Universt\"at Mainz, Germany\\
$^3$ Laboratori Nazionali di Frascati dell'INFN, Frascati, Italy\\

The importance of continuous and close collaboration between the experimental
and theoretical groups is crucial in the quest for
precision in hadronic physics.
This is the reason why the  
Working Group on ``Radiative Corrections and Monte Carlo Generators for Low Energies'' (Radio MonteCarLow)  was formed a few years ago bringing together experts (theorists and experimentalists) working in the field of low-energy $e^+e^−$ physics and partly also the $\tau$ community.
Its main motivation  was to understand the status and the precision of the Monte Carlo generators used to analyse the hadronic cross section measurements
obtained as well with energy scans as with radiative return, to determine luminosities, and whatever possible to perform tuned comparisons, {\it i.e.}
comparisons of MC generators with a common set of input parameters and experimental cuts. This  main effort was summarized in a report published in 2010~\cite{Actis:2010gg}.
During the years the WG structure has been enriched of more physics items 
and now it includes seven subgroups: Luminosity, R-measurement, ISR, 
Hadronic VP $g-2$ and Delta alpha, gamma-gamma physics, FSR models, tau. 

During the workshop the last achievements of each subgroups have been presented. A particular emphasis has been put on the 
recent evaluations of the hadronic contributions to the $g-2$ of the muon.
For the hadronic VP contribution the necessity to have a database with an easy general access of hadronic cross section measurements with a clear indication of the treatment of Radiative Corrections and systematic errors was discussed.
For the hadronic Light-by-Light contribution different theoretical models were presented and the possibility to constrain some of them by data was discussed.
Finally   a proposal for a white book on meson transition form factors was presented.\\

All the information on the Radio MonteCarLow WG can be found at the web page:
\begin{center}
\url{http://www.lnf.infn.it/wg/sighad/} 
\end{center}

\newpage

\section{Short summaries of the talks}

\subsection{Review of R measurements and perspectives at BES-III}
\addtocontents{toc}{\hspace{2cm}{\sl A.~Denig}\par}

\vspace{5mm}

A.~Denig

\vspace{5mm}

\noindent
Institut f\"ur Kernphysik, Johannes Gutenberg Universt\"at Mainz, Germany\\

\vspace{5mm}

The R ratio is defined as the ratio of the total hadronic cross section in electron-positron annihilation
normalized to the di-muon cross section:
\begin{equation}
R=\frac{\sigma(e^+e^- \to {\rm hadrons})}{\sigma(e^+e^- \to \mu^+\mu^-)}.
\end{equation}
Precise measurements of the R ratio are closely related to the development of the Standard Model of particle physics. Furthermore, many fundamental parameters of QCD can be derived from a precise knowledge of R, such as for instance the strong coupling constant and the charm and bottom quark masses. \\
A significant experimental progress regarding $R$ was achieved in the past years at energies below
approximately 12 GeV. This progress was phenomenologically motivated by the impact of R measurements on two precision quantities: (i) the anomalous magnetic
moment of the muon $(g-2)_\mu$ and (ii) the value of the electromagnetic fine structure constant at the Z pole $\alpha_{\rm QED}(M_Z^2)$. Both quantities play an important role in precision tests of the Standard Model and are limited by hadronic vacuum polarization
effects. The hadronic contributions can be derived via dispersion integrals, using experimental R data as input.
For the phenomenological evaluations of these hadronic contributions
different energy ranges of hadronic cross section data are required.
While the energy range below 2 GeV is almost sufficient for $(g-2)_\mu$, data at higher
energies -- especially above 2 GeV -- is needed for improving the hadronic contribution to
$\alpha_{\rm QED}(M_Z^2)$ \cite{davier} \cite{jegerlehner}.
\\
\\
Besides the conventional energy scan, more recently a new and complementary method has been established at so-called particle factories, which are designed to run at fixed center-of-mass energies. In the so-called Radiative Return technique, events are considered in which the initial beam leptons radiate a high-energetic photon, lowering in such a way the available energy for the production of hadrons in the final state. The method has
been applied at essentially all modern $e^+e^-$ accelerators and many precision measurements of exclusive hadronic final states have been performed \cite{review_isr} \cite{radiomc}.
\\
\\
So far, the world's best measurement of the inclusive R ratio above 2 GeV has been measured by the BES collaboration. After a pre-study at BES-I \cite{bes1}, two scanning campaigns with 6 \cite{bes2a}
and later 85 \cite{bes2b} continuum points between 2.0 GeV and 5.0 GeV have been performed by BES-II.  The individual energy points
had a statistical accuracy of 3-5 \% and a systematic accuracy of 5-8 \%. Later, few
scan points \cite{bes2c} \cite{bes2d} close to the charmonia resonances have been taken in addition. The systematic accuracy
has been improved by almost a factor 2 in these recent results.
While the BES measurements were inclusive measurements of R, exclusive measurements of open charm channels have been obtained via ISR by the BELLE \cite{belle_a} and BABAR \cite{babar_a} collaborations. Furthermore, BABAR \cite{babar_b} has also covered the bottomonium region via a dedicated energy scan.
\\
\\
At the BES-III experiment a new energy scan is planned between 2.0 and 4.6 GeV.  Thanks to
the good performance of the BEPC-II accelerator and the BES-III detector it is foreseen to improve significantly
upon the existing BES-II measurement. The inclusive R ratio will be measured with about 1 \% statistical and 3 \% systematic accuracy. The measurement will be performed in 3 phases:
\begin{itemize}

\item{{\bf Phase 1:} Low statistics run between 2 - 4.6 GeV with 1 \% statistical uncertainty per scan point; {\bf motivation: $\alpha_{\rm QED}(M_Z^2)$}}

\item{{\bf Phase 2:} High statistics run between 2 - 3 GeV; {\bf motivation: nucleon and hyperon electromagnetic form factors}}

\item{{\bf Phase 3:} Fine scanning in the charmonium region, exclusive measurement of charmonium pairs; {\bf motivation: extraction of charm quark mass and charmonium spectroscopy}}

\end{itemize}

\newpage

\subsection{Status of the CMD-3 Experiment at VEPP-2000}
\addtocontents{toc}{\hspace{2cm}{\sl S.~Eidelman}\par}

\vspace{5mm}

S.~Eidelman

\vspace{5mm}

\noindent
      Budker Institute of Nuclear Physics SB RAS and  \\
Novosibirsk State University, Novosibirsk, Russia \\

\vspace{5mm}

Since 2010 a new $e^+e^-$ collider VEPP-2000~\cite{vepp1,vepp2} has been
taking data with
two detectors, SND~\cite{snd} and CMD-3~\cite{cmd3}.
During this period the following integrated
luminosity was collected: 3.1 pb$^{-1}$ at the $\phi$, 33 pb$^{-1}$ from
the $\phi$ to 2 GeV and 5.2 pb$^{-1}$ below the $\phi$. \\

The maximum luminosity achieved by now is $2 \cdot 10^{31}$ cm$^{-1}$s$^{-1}$
at 1.7-1.8 GeV and it falls much slower with decreasing energy than
before. At high energies the luminosity is limited by a deficit of positrons
and maximum energy of the booster, which is 900 MeV.
A long shutdown for 1-1.5 years to increase the booster energy to 1 GeV
and commission the new injection complex to reach $10^{32}$ cm$^{-1}$s$^{-1}$
is expected to begin this autumn. \\

Analysis is in progress for various final states, in particular
$e^+e^- \to \pi^+\pi^-,~K^+K^-,~\pi^+\pi^-\pi^0$,
$2\pi^+2\pi^-,~\pi^+\pi^-2\pi^0,~K^+K^-\pi^+\pi^-,~3\pi^+3\pi^-,
~2\pi^+2\pi^-2\pi^0,~p\bar{p}$.   \\

In a scan from 1500 to 2000 MeV with a $\sqrt{s}=25$~MeV step
and a finer scan of the near-$N\bar{N}$ threshold an integrated
luminosity of 22~pb$^{-1}$ was collected and used to study the
process $e^+e^- \to 3\pi^+3\pi^-$~\cite{6pi}.
About $8 \times 10^3$ five- and six-track events were selected
(5069 and 2887 events, respectively) and a cross section of the process
was determined as a function of energy. Its values and energy dependence
are consistent with those of BaBar~\cite{babar} and confirm a dip
around 1.9 GeV. It is worth mentioning that the precision reached
at this relatively small integrated luminosity is similar to that of
BaBar. \\

One can hope that in the close future a lot of new precise information
will be available that will allow one to perform new, more precise
calculations of the hadronic contributions to $\amu$, the muon
anomalous magnetic moment.

\newpage

\subsection{Improvement of the Lund area law hadronic generator LUARLW}
\addtocontents{toc}{\hspace{2cm}{\sl H.~Hu}\par}

\vspace{5mm}

H.~Hu

\vspace{5mm}

\noindent
Institute of High Energy Physics, Beijing, China\\

\vspace{5mm}

The Born cross section ($R$ value) of the inclusive hadronic production in $e^+e^-$ annihilation is an important parameter in the Standard Model. $R$ values under $5$ GeV are obtained by experiments. In measurement the detecting efficiency of the hadronic events is determined by the Monte Carlo generators and the detector simulations, and among the total systematic errors, the contribution from hadronic model is dominant.

The pivotal part in a hadronic generator is the hadronization scheme. The famous inclusive generator JETSET uses the Lund string fragmentation function~\cite{JETSET}, and the LUARLW adopts the Lund string fragmentation area law as its hadronization model~\cite{bo,bohu}. LUARLW borrows several subroutines in JETSET to treat some subprocesses except for hadronization, such as unstable particles decay. The earlier version of LUARLW was developed about 15 years ago, and used in the $R$ value measurements with BESII~\cite{Rprl2002,Rplb2008}. The old LUARLW only simulates the quark string fragmentation and the production of continuous hadronic states~\cite{hnpnp2011}.
The simulation scheme of the initial state radiation is described in references~\cite{SLAC86,SLAC90}. The typical errors of the hadronic efficiency of LUARLW are about 2.5$\%$ then.

BESIII plans to perform the fine scan of the line-shape of hadronic production structure from 2 to 4.6 GeV at about 100 energy points with large integrated luminosity and small energy step lengths, and measure the $R$ value and resonant parameters of charmonium with $J^{PC}=1^{--}$ more precisely, the error decrease to about $3\%$~\cite{BESIIIyb}. In order to simulate all hadronic states with a consistent generator and decrease the systematic error of hadronic efficiency, LUARLW has been improved significantly, some new production and decay channels are added. Now LUARLW can simulates following hadronic processes:\\
\[
e^+e^-\Rightarrow\gamma^\ast\Rightarrow \left\{\begin{array}{ll}

V~(\rho,\omega,\phi)\\

q\bar{q}\Rightarrow {\rm string}\Rightarrow{\rm hadrons}&\\

gq\bar{q} \Rightarrow {\rm string~+~string}\Rightarrow {\rm hadrons} & \\

\end{array}
\right.
\]
\[
e^+e^-\Rightarrow\gamma^\ast\Rightarrow
J/\psi\Rightarrow\left\{\begin{array}{ll}

\gamma^\ast\Rightarrow e^+e^-,~\mu^+\mu^- & \\

\gamma^\ast\Rightarrow q\bar{q}\Rightarrow {\rm string}\Rightarrow{\rm hadrons}&\\

ggg \Rightarrow {\rm
string~+~string~+~string~}\Rightarrow {\rm hadrons} & \\

\gamma gg \Rightarrow {\rm
string~+~string~}\Rightarrow {\rm hadrons} & \\

\gamma\eta_c \Rightarrow gg \Rightarrow {\rm
string~+~string~}\Rightarrow {\rm hadrons} &\\

{\rm exclusive~radiative~decay~channels}

\end{array}
\right.
\]
\[
e^+e^-\Rightarrow\gamma^\ast\Rightarrow
\psi(3686)\Rightarrow\left\{\begin{array}{ll}

\gamma^\ast\Rightarrow e^+e^-,~\mu^+\mu^-,~\tau^+\tau^- & \\

\gamma^\ast\Rightarrow q\bar{q}\Rightarrow {\rm string}\Rightarrow{\rm hadrons}&\\

ggg \Rightarrow {\rm
string~+~string~+~string~}\Rightarrow {\rm hadrons} & \\

\gamma gg \Rightarrow {\rm
string~+~string~}\Rightarrow {\rm hadrons} & \\

\gamma\eta_c \Rightarrow gg \Rightarrow {\rm
string~+~string~}\Rightarrow {\rm hadrons} &\\

\pi\pi J/\psi,~\eta J/\psi,~\pi^0 J/\psi & \\

{\rm exclusive~radiative~decay~channels}

\end{array}
\right.
\]
\[
e^+e^-\Rightarrow\gamma^\ast\Rightarrow
\psi(3770)\Rightarrow\left\{\begin{array}{ll}

\gamma^\ast\Rightarrow e^+e^-,~\mu^+\mu^-,~\tau^+\tau^- & \\

D\bar{D}& \\

\gamma^\ast\Rightarrow q\bar{q}\Rightarrow {\rm string}\Rightarrow{\rm hadrons}&\\

ggg \Rightarrow {\rm
string~+~string~+~string~}\Rightarrow {\rm hadrons} & \\

\gamma gg \Rightarrow {\rm
string~+~string~}\Rightarrow {\rm hadrons} & \\

\pi\pi J/\psi,~\eta J/\psi,~\pi^0 J/\psi & \\

{\rm exclusive~radiative~decay~channels} & \\

\end{array}
\right.
\]
\[
e^+e^-\Rightarrow\gamma^\ast\Rightarrow \left\{\begin{array}{ll}

\psi(4040)\Rightarrow  D\bar{D},D^\ast\bar{D}^\ast, D\bar{D}^\ast, D^\ast\bar{D}, D_s\bar{D}_s, {\rm other~decay~modes}& \\

\psi(4160)\Rightarrow  D\bar{D},D^\ast\bar{D}^\ast, D\bar{D}^\ast, D^\ast\bar{D}, D_s\bar{D}_s, D_s\bar{D}_s^\ast, {\rm other~decay~modes}& \\

\psi(4415)\Rightarrow  D\bar{D},D^\ast\bar{D}^\ast, D\bar{D}^\ast, D^\ast\bar{D}, D_s\bar{D}_s, D_s\bar{D}_s^\ast, D_s^\ast\bar{D}_s^\ast, {\rm other~decay~modes}& \\

X(4260), X(4360)\Rightarrow {\rm possible~decay~modes}
\end{array}
\right.
\]

There are many phenomenological parameters in LUARLW. The most important and sensitive one for determining hadronic efficiency are those related to the multiplicity, polar angle and momentum distributions of the preliminary hadrons. The ratios of the different hadronic production channels in string fragmentation and the ratios of mesons and baryons with different quantum numbers are set by the scheme used in JETSET~\cite{JETSET}, but some default values are replaced by the one tuned in reference~\cite{DELPHI} or set by the values cited in PDG~\cite{PDG}. In the parameters tuning the effort of only using unique group of parameters to fit all data samples in whole BEPCII energy region are tried. The most of the final states distributions related to events criteria at detector level agree with data well, and more fine parameters tuning are continuing. The goal of LUARLW improvement and parameters tuning is the error of hadronic efficiency could be decreased to about $2\%$.

\newpage

\subsection{Measurement of Hadronic Cross Sections Using Initial State Radiation}
\addtocontents{toc}{\hspace{2cm}{\sl B.~Kloss}\par}

\vspace{5mm}

 A.~Denig and B.~Kloss

\vspace{5mm}

\noindent
Institut f\"ur Kernphysik, Johannes Gutenberg Universt\"at Mainz, Germany\\

\vspace{5mm}

\noindent Cross sections of the form $e^+e^-\rightarrow hadrons$ are an important input for the standard model prediction of the hardonic contribution to the anomalous magnetic moment of the muon $a_\mu$ \cite{Jegerlehner}. The hadronic contribution caused by vacuum polarization can be calculated with a dispersion integral
\begin{equation}
	a_\mu^{hadr} \cong \frac{1}{4\pi^3}\int_{4m_\pi^2}^\infty K(s)\sigma (e^+e^-\rightarrow hadrons)ds
\end{equation}
where $K(s)\propto\frac{1}{s}$ is the so called Kernel Function. The experimental uncertainty in these hadronic cross sections limits the standard model prediction completely. \\
\\
The largest contribution to the absolute value of $a_\mu^{hard}$ comes from cross sections at an energy below 1 GeV, i.e. the $\pi^+\pi^-$ cross section. This one has been measured with high precision at the BaBar, KLOE and CMD2 experiments \cite{BaBar,KLOE,CMD2}. For the error $\Delta a_\mu^{hadr}$ contributions between 1 and 2 GeV get more important, which means the $\pi^+\pi^-\pi^0$, $\pi^+\pi^-\pi^0\pi^0$ and $\pi^+\pi^-\pi^+\pi^-$ final states. Our goal is to measure these cross sections at the BaBar and BES-III experiments \cite{babar,design and construction of BES-III} with a very high precision.\\
\\
Therefor we want to use the technique of Initial State Radiation \cite{ISR}. If a photon is emitted in the initial state the center of mass energy is lowered by the energy of the emitted photon. So measurements of cross sections at different energies are possible although the collider has a fixed cms energy. By measuring the ISR cross section it is then possible to extract the non-radiative cross section which is the input for the dispersion integral via
\begin{equation}
	\frac{d\sigma_{ISR}(M_{hadrons})}{dM_{hadrons}} = \frac{2M_{hadrons}}{s}\cdot W(s,x,\theta_\gamma) \cdot \sigma(M_{hadrons})
\end{equation}
where $M_{hadrons}$ is the invariant mass of the hadronic system and $W$ the so called Radiator Function which gives the probability that the ISR photon is emitted with a specific energy fraction $x$ and angle $\theta_\gamma$. For the Monte-Carlo prediction we are using the ISR generator PHOKHARA 7.0 \cite{PHOKHARA1,PHOKHARA2,PHOKHARA3}.\\
\\
The cms energy $s$ at the BaBar experiment is 10.58 GeV where the $\pi^+\pi^-\pi^0\pi^0$ and the $\pi^+\pi^-\pi^+\pi^-$ final states are under study in Mainz \cite{4pi,2pi2pi0}. Whereas $s = 3.773$~GeV at the BES-III experiment where currently the $\pi^+\pi^-$, $\pi^+\pi^-\pi^0$, $\pi^+\pi^-\pi^0\pi^0$ cross sections are under investigation. We hope that we are able to make a contribution to the precise measurement of these hadronic cross sections.

\newpage

\subsection{Precise Charm and Bottom Quark Masses}
\addtocontents{toc}{\hspace{2cm}{\sl J.H.~K\"uhn}\par}

\vspace{5mm}

J.H.~K\"uhn

\vspace{5mm}

\noindent
Institut f\"ur Theoretische Teilchenphysik, Karlsruhe Institut f\"ur Technologie,\\
76128 Karlsruhe, Germany

\vspace{5mm}

{\bf 1. Method\\}

Exploiting the analyticity of the vacuum polarization function
$\Pi(q^2)$ around $q^2=0$ and using dispersion
relations, the derivatives at $q^2=0$ can be expressed as weighted
integrals over the imaginary part of $\Pi(q^2)$, which in turn is given
by the cross section for electron-positron annihilation into
hadrons. Let us denote the normalised cross section for heavy quark
production as $R_Q(s)\equiv \sigma_Q(s)/\sigma_{\rm point}(s)$. The
moments of $R_Q$, defined as
\begin{equation}
  {\cal M}^{\rm exp}_n \equiv \int \frac{{\rm d}s}{s^{n+1}} R_Q(s)
  \,,
  \label{eq:Mexp}
\end{equation}
can be directly related to the perturbatively calculated Taylor
coefficients. In total one thus obtains the $\overline{MS}$ quark mass in
terms of experimentally weighted integrals of $R_Q$ and the perturbatively
calculable coefficients $\bar C_n$,
\begin{equation}
  m_Q(\mu) = \frac{1}{2}
  \left( \frac{9 Q_Q^2\bar{C}_n}{4 {\cal M}_n^{\rm exp}}\right)^{1/(2n)}
  \,.
  \label{eq:m_Q}
\end{equation}

This strategy has been suggested originally in
Ref.~\cite{Shifman:1978bx} and applied to a
precise charm and bottom mass determination in Ref.~\cite{Kuhn:2001dm}
once the
three-loop results had become available. A significantly improved reanalysis
based on four-loop moments as obtained in Refs.~%
\cite{Chetyrkin:2006xg,Boughezal:2006px,Maier:2008he,Maier:2009fz,Sturm:2008eb,%
Kiyo:2009gb}
and with new data has been performed
in Ref.~\cite{Kuhn:2007vp}, additional
updates and improvements from new data and the precise analytical evaluation of the
perturbative moments can be
found in Refs.~\cite{Chetyrkin:2009fv,Chetyrkin:2010ic}.
For the extraction of $R_Q$ from the data the issue of singlet
contributions and secondary radiation of heavy quarks has been discussed
in some detail in Ref.~\cite{Kuhn:2007vp}.
Furthermore, the potential influence of a
non-vanishing gluon condensate
$\langle\frac{\alpha_s}{\pi}GG\rangle=0.006\pm0.012\,{\rm GeV}^4$
on the charm mass determination
has been analysed \cite{Kuhn:2007vp,Chetyrkin:2010ic}
and found to be small.\\

{\bf 2. Results\\}

Let us now present the experimental results for the moments, first for
charm, later for bottom. For charm the integration region is split into
one covering the narrow resonances $J/\psi$ and $\psi'$, the ``threshold
region'' between $2 m_D$ and 4.8~GeV and the perturbative continuum
above. Note that we assume the validity of pQCD in this region with high
precision, an assumption that is well consistent with present
measurements but for the moment remains an assumption, to be
verified e.g. by future BESS experiments (for charm) and Belle (for bottom).

  \begin{table}[t]
  \begin{center}
{
\begin{tabular}{l|lll|l||l}
\hline
$n$ & ${\cal M}_n^{\rm res}$
& ${\cal M}_n^{\rm thresh}$
& ${\cal M}_n^{\rm cont}$
& ${\cal M}_n^{\rm exp}$
& ${\cal M}_n^{\rm np}$(NLO)
\\
& $\times 10^{(n-1)}$
& $\times 10^{(n-1)}$
& $\times 10^{(n-1)}$
& $\times 10^{(n-1)}$
& $\times 10^{(n-1)}$
\\
\hline
$1$&$  0.1201(25)$ &$  0.0318(15)$ &$  0.0646(11)$ &$  0.2166(31)$
 &$-0.0002(5)$ \\
$2$&$  0.1176(25)$ &$  0.0178(8)$ &$  0.0144(3)$ &$  0.1497(27)$
 &$-0.0005(10)$ \\
$3$&$  0.1169(26)$ &$  0.0101(5)$ &$  0.0042(1)$ &$  0.1312(27)$
 &$-0.0008(16)$ \\
$4$&$  0.1177(27)$ &$  0.0058(3)$ &$  0.0014(0)$ &$  0.1249(27)$
 &$-0.0013(25)$ \\
\hline
\end{tabular}
\caption{Experimental moments
  in $(\mbox{GeV})^{-2n}$ as defined in
  Eq.~(\ref{eq:Mexp}), separated according to the contributions from
  the narrow resonances,
  the charm threshold region and the continuum region
  above $\sqrt{s}=4.8$~GeV. In the last column the NLO contribution from the
  gluon condensate is shown.}
}
\label{tab:contglu}
\end{center}
  \end{table}

  \begin{table}[t]
  \begin{center}
{
\begin{tabular}{l|l|llll||l}
\hline
$n$ & $m_c(3~\mbox{GeV}) $ & exp & $\alpha_s$ & $\mu$
& ${\rm np}_{\rm NLO}$ & total\\
 \hline
1& 0.986& 0.009&  0.009&  0.002& 0.001& 0.013\\
2& 0.975& 0.006&  0.014&  0.005& 0.002& 0.016\\
3& 0.975& 0.005&  0.015&  0.007& 0.003& 0.017\\
4& 0.999& 0.003&  0.009&  0.031& 0.003& 0.032\\
\hline
\end{tabular}
  \caption{Results for $m_c(3\, {\rm GeV} )$ in ${\rm GeV}$. The errors
      are from experiment, $\alpha_s$, variation of $\mu$ and the
     gluon condensate.}
    \label{tab:charmTab}}
    \end{center}
\  \end{table}
The results for the moments from one to four and the error budget are
listed in Table~1,
those for the quark mass in Table~2.
The moment with $n=1$ is most robust from the theory side, as
evident from the relatively smaller coefficient in the perturbative
series.
In view of the smallest sensitivity to
$\alpha_s$ and to the choice of the renormalisation scale $\mu$ we adopt
the value as derived from $n=1$ as our final result:
\begin{equation}
m_c(3\,{\rm GeV})=986(13)\,{\rm MeV}.
\end{equation}

Tables~1 and 2 also
illustrate the path to a further reduction of the error.
For $n=1$ important contributions arise
from all three regions. Improved determinations of $\Gamma_e(J/\psi)$
would reduce the errors of all three moments. Improved measurements of
$R_Q$ in the threshold region and at 4.8~GeV would have a strong impact
on $n=1$ and strengthen our confidence in the validity of pQCD close to,
but above 4.8~GeV. Another interesting option would be a simultaneous fit
to all three moments, taking the proper experimental correlations into
account.

Similar statements can be made for the determination of the bottom quark
mass. A recent measurement of the cross section in the threshold
region between 10.6~GeV and 11.2~GeV was employed in
Ref.~\cite{Chetyrkin:2009fv} and has lead to a significant reduction
of the experimental error on $m_b$. Still, additional measurements in
the region around and above 11~GeV would be welcome in order to confirm
the validity of perturbative QCD relatively close to threshold.
The result for the second moment has been adopted as our final answer
\begin{equation}
m_b(10\,{\rm GeV}) = 3610(16)\,{\rm MeV}
\end{equation}
and corresponds to $m_b(m_b)=3610(16)\,{\rm MeV}$.

\newpage

\subsection{Measurement of the time-like $\eta$ and $\pi^0$ transition from factors using initial state radiation}
\addtocontents{toc}{\hspace{2cm}{\sl A.~Kupsc}\par}

\vspace{5mm}

\noindent
 A.~Kupsc and J. Pettersson

\vspace{5mm}

\noindent
Department of Physics and Astronomy, Uppsala University, Sweden\\

\vspace{5mm}

Knowledge  of  meson  transition  form  factors is  needed  to  reduce
uncertainty  of  the  Standard  Model  value for  the  muon  anomalous
magnetic moment,  since double  off--shell form factors  are necessary
input   for   the   calculation   of   the   hadronic   light-by-light
contribution~\cite{Hayakawa:1997rq,Bijnens:2007pz}.     The    physics
issues  and  prospects  of  new  high-precision  measurements  of  the
$\pi^0$,  $\eta$  and  $\pi^0$   meson  transition  form  factors  are
discussed in Refs.~\cite{Landsberg:1986fd,Czerwinski:2012ry}.

Measurement of  the $e^+e^-\to P\gamma$ reaction  cross section (where
$P=\pi^0,\eta$) as a function of the center-of-mass energy $\sqrt{s}$,
gives  access to the  transition form  factor $F_P(s,0)$  in time-like
region:
\begin{equation}
\sigma(e^+e^-\to P\gamma)=4\pi\alpha\Gamma_{\gamma\gamma}
\left(\frac{s-m_P^2}{s\, m_P}\right)^3|F_P(s,0)|^2. \label{eq:ee-Pg}
\end{equation}
Presently the  only information on  these reactions for  $\sqrt{s}< 1$
GeV      comes       from      CMD-2~\cite{Akhmetshin:2004gw}      and
SND~\cite{Achasov:2003ed,Achasov:2007kw}   experiments   with  limited
statistics.

We  would like  to address  the question  whether one  could  use high
integrated   luminosities  from  $\phi$   (KLOE)  \cite{Bossi:2008aa},
$J/\psi$  (BESIII) \cite{Li:2011ve}  or  B factories  and the  initial
state   radiation  method   in   order  to   extract  the   $e^+e^−\to
\pi^0\gamma/\eta\gamma$ cross section at  $\sqrt{s} < 1$ GeV.  Ideally
a goal of the studies would be to provide constrains for the low $s$
behavior of  the form  factors which is  approximated in terms  of the
slope: $F_P(s,0)=1+as$.

In the  first step of  the feasibility studies  the Born terms  of the
$e^+e^-\to\pi^0\gamma\gamma_{ISR}/\eta\gamma\gamma_{ISR}$     reactions
were   included    in   the   Phokara   7.0    ISR   event   generator
\cite{Rodrigo:2001kf,Czyz:2002np}.    This  will  allow   to  simulate
acceptance and  the expected event yields  and compare to  the size of
the expected background. Another  problem is related to the extraction
of the transition form factor from the measured cross section.  Here a
method have  to deal with  a combinatorial ambiguity  and interference
between  the  ISR and  FSR  photons.   Preliminary  studies show  that
a satisfactory selection  could be obtained  by tagging the  ISR photon
as the  one with maximum  likelihood given  by the following
radiator distribution \cite{Druzhinin:2011qd} where electron masses are
take to be small,  depending on the energy and the scattering angle:

\begin{equation}
w_{0}(\theta,x)=\frac{\alpha}{\pi}\left[\frac{x}{2\tan(\theta)^{2}}+\frac{1-x}{x\sin(\theta)^{2}}\right]
\end{equation}

Here $x$ is the fraction of beam momenta carried away by the ISR photon.

\newpage

\subsection{Final state emission radiative corrections to the process $e^+e^- \to \pi^+\pi^-(\gamma)$. Contribution to muon anomalous magnetic moment}

\addtocontents{toc}{\hspace{2cm}{\sl E.~A.~Kuraev}\par}

\vspace{5mm}

\noindent
A.~I.~Ahmadov$^{1,3}$, E.~A.~Kuraev$^{1}$, M.~K.~Volkov$^{1}$, O.~O.~Voskresenskaja$^{2}$ and E.~V.~Zemlyanaya$^{2}$

\vspace{5mm}

\noindent
$^{1}$ Bogoliubov Laboratory of Theoretical Physics, Joint Institute for Nuclear Research, Dubna, 141980 Russia, \\
$^{2}$ Laboratory of Information Technologies, Joint Institute for Nuclear Research, Dubna, 141980 Russia,  \\
$^{3}$ Institute of Physics, Azerbaijan, National Academy of Sciences, Baku, Azerbaijan

\vspace{5mm}

Analytic calculation of contribution to anomalous magnetic moment of muon from the channels of annihilation of electron-positron pair to the pair of charged pi-meson with radiative correction connected with final state, as well as corrections to the lowest order kernel are presented. The result with the point-like pi-meson assumptions is 

$a_{point}=a_{point}^{(1)}+\Delta a_{point}$,\,\, $a_{point}^{(1)}=7.0866 \cdot 10^{-9}; \,\, \Delta a_{point}= -2.4 \cdot 10^{-10} .$

Taking into account the pion form factor in frames of Nambu-Jona-Lasinio (NJL) approach leads to

$a_{NJL}=a_{NJL}^{(1)}+\Delta a_{NJL}$,\,\, $a_{NJL}^{(1)}= 5.48\cdot 10^{-8}; \,\, \Delta a_{NJL}=-3.43\cdot 10^{-9}$.

The result, obtained in point-like approximation about an order of magnitude lower than one measured in experiment \cite{Davier}

$a_\mu\approx 6.974\cdot 10^{-8}$.

Conversion of a virtual photon to $\pi^+\pi^-(\gamma)$ state in time-like region is realized through the intermediate state with vector meson $\rho(775), \omega(782), \phi(1020), \rho'(1450)$ with the following decay to the two pion state. Main contribution arise from $\rho(775), \omega(782)$ meson states.

\newpage

\subsection{Estimated calculation of the hadronic light-by-light contribution to the $(g-2)$ of the muon}
\addtocontents{toc}{\hspace{2cm}{\sl P.~Masjuan}\par}

\vspace{5mm}

\noindent
P.~Masjuan

\vspace{5mm}

\noindent

Institut f\"ur Kernphysik, Johannes Gutenberg Universt\"at Mainz, Germany\\

\vspace{5mm}

The anomalous magnetic moment of the muon is one of the most accurately measured quantities in particle physics. Any deviation from its prediction in the Standard Model of particle physics is a very promising signal of new physics. The present world average experimental value is given by $\amu^{EXP}=116 592 08.9(6.3)\times10^{-10}$ \cite{Bennett:2004pv,Bennett:2006fi}. A proposal to measure the muon $(g-2)_{\mu}$ to a precision of $1.6 \times 10^{-10}$ has recently been submitted to FNAL \cite{Carey:2009zzb}.

The QED contributions to $\amu$ are very well known (up to the fifth order ${\cal O}(\alpha_{em}^5)$), giving the result $11658471.885(4)\times10^{-10}$ \cite{Aoyama:2012wk}, followed by the electroweak one ($15.4(2)\times10^{-10}$ \cite{Czarnecki:2002nt}). The main uncertainties at present originate from the hadronic vacuum polarization (HVP) ($692.3(4.2)\times10^{-10}$ \cite{Davier:2010nc} at leading order and  $-9.84(7)\times10^{-10}$ \cite{Hagiwara:2011af} at higher orders) as well as hadronic light-by-light scattering corrections (HLBL) ($11.6(4.0)\times10^{-10}$  \cite{Jegerlehner:2009ry}). The existing discrepancy between the experimental value for $(g-2)_{\mu}$ and its Standard Model prediction stands at about $3\sigma$.

The HLBL cannot be directly related to any measurable cross section, and requires the knowledge of QCD contributions at all energy scales. Since this is not known yet, one needs to rely on hadronic models to compute it. Such models introduce some systematic errors which are difficult to quantify.

We provided~\cite{Masjuan:2012qn} an estimated calculation with a judicious error estimate for the HLBL scattering based on a duality argument between the hadronic degrees of freedom and the well-known quark loop contribution. Such a duality estimate can be obtained by invoking the large-$N_c$ of QCD \cite{tHooft:1973jz,Witten:1979kh} where a quark-hadron duality is accounted for considering that hadronic amplitudes are described by an infinite set of non-interacting and non-decaying resonances.

Using the large-$N_c$ counting and also the chiral counting, it was proposed in \cite{deRafael:1993za} to split the full HLBL into a set of different contributions where the numerically dominant contribution arises from the pseudo-scalar exchange diagram. Calculations carried out in the large-$N_c$ limit demand an infinite set of resonances for computing any quantity. As such sum is not known in practice, one ends up truncating the spectral function in a resonance saturation scheme.

It was pointed out in Refs.~\cite{Masjuan:2007ay,Masjuan:2008fr} that, in the large-$N_c$ framework, the resonance saturation can be understood from the mathematical theory of Pade Approximants (PA) to meromorphic functions, where one can compute the desired quantities in a model-independent way and even be able to ascribe a systematic error to the approach \cite{Masjuan:2009wy}.

For our evaluation we will project the hadronic content of the quark loop onto the dominant hadronic piece of the HLBL~\cite{Knecht:2001qf}. We will use, instead of a hadronic model for the transition Form Factors (FF), a sequence of rational approximants  \cite{Masjuan:2007ay,Masjuan:2008fr} build up from the low-energy expansion of the pion FF obtained in \cite{Masjuan:2012wy} after a fit to the experimental data, to minimize a model dependence~(see \cite{Masjuan:2012wy,Masjuan:2008fv,Masjuan:2012qn} for details). The offshellness effects are considered as in Refs.~\cite{Melnikov:2003xd,Jegerlehner:2009ry}. We use the half-width rule to account for the masses in the Large-$N_c$ limit~\cite{Masjuan:2012gc,Masjuan:2012sk}.

Such model yields the average momenta running into the dominant piece of the HLBL. By the duality argument, such momenta corresponds to the one in the quark-loop diagram. Lattice QCD could be able to calculate the momentum dependent quark masses for such quark-loop. As at present this is not yet fully feasible for physical pion masses, our proposal is to use the dressed-quark mass function $M(Q)$ in the chiral limit computed within the Dyson-Schwinger equation (DSE) framework which provides us with the desired momentum dependent quark-mass function \cite{Bhagwat:2007vx}. We add a $10\%$ relative error as an estimate for the extrapolation from the chiral limit to the physical light quark mass regime. With this mass, one can use the well-known formulae for the $a_{\mu}^{HLBL}$ for spin$-1/2$ fermions computed in \cite{Laporta:1992pa}.

To quote a final number for $\amu^{HLBL}$ implies to consider several sources of error (inputs for the FF,  $10\%$ error due to the departure from the chiral limit in the quark-mass momentum-dependent function, a $5\%$ a systematic error from the PA sequence used) and leads to our ballpark estimation:

\begin{equation}
\amu^{HLBL}=[8.2(1) \div 12.6(2)]\times 10^{-10}\, ,
\end{equation}
\noindent
where the error in parenthesis is the input errors and the two numbers represent the range due to the departure from the chiral limit considered in our computations.

\newpage

\subsection{The Why's and How's of covariance matrices in the KLOE ISR analyses}
\addtocontents{toc}{\hspace{2cm}{\sl S.~E.~M\"uller}\par}

\vspace{5mm}

S.~E.~M\"uller

\vspace{5mm}

\noindent

Institute of Radiation Physics, Helmholtz-Zentrum Dresden-Rossendorf, Dresden, Germany\\

\vspace{5mm}

The data points from the KLOE ISR analyses~\cite{Aloisio:2004bu,Ambrosino:2008aa,Ambrosino:2010bv2,Babusci:2012rp} on the cross section for the process $(e^+ e^- \to \pi\pi)$ come with non-diagonal covariance matrices (see~\cite{KLOEurl}). In the following, I will discuss the implications of these covariance matrices on the uncertainty of the hadronic contribution to the muon anomaly.

Let's recall that the uncertainty of a function $f=f(x_1,x_2,\dots,x_n)$ which depends on a set of (correlated) observables $x_1,x_2,\dots,x_n$ is constructed via

\begin{equation}
\sigma_f^2 = \left(\frac{\partial f}{\partial x_1},\frac{\partial f}{\partial x_2},\dots,\frac{\partial f}{\partial x_n}  \right) \left(
\begin{array}{ccccc}
\sigma_1^2 & \sigma_{12} & \sigma_{13} & \ldots & \sigma_{1n}\\
\sigma_{21} & \sigma_2^2 & \sigma_{23} & \ldots & \sigma_{2n}\\
\sigma_{31} & \sigma_{32} & \sigma_3^2 & \ldots & \sigma_{3n}\\
\vdots & \vdots & \vdots & \ddots & \vdots \\
\sigma_{n1} & \sigma_{n2} & \sigma_{n3} & \ldots & \sigma_{n}^2
\end{array}
\right) \left(
\begin{array}{c}
{\partial f}/{\partial x_1} \\
{\partial f}/{\partial x_2} \\
\vdots \\
{\partial f}/{\partial x_n}
\end{array}
\right) \, ,
\label{eq:1}
\end{equation}

where the diagonal elements of the matrix contain the squared uncertainties (variances)  $\sigma_i^2$ and off-diagonal elements contain the covariances $\sigma_{ij}=\rho\sigma_i \sigma_j$ with the correlation coefficient $\rho$ (bound by $-1 \le \rho \le +1$). If the observables are uncorrelated, the $\rho$'s are zero and the covariance matrix is diagonal. In this case, one recovers the familiar formula

\begin{equation}
\sigma_f^2 = \left(\frac{\partial f}{\partial x_1}\right)^2\sigma_1^2 +\left(\frac{\partial f}{\partial x_2}\right)^2\sigma_2^2 +\dots + \left(\frac{\partial f}{\partial x_n}\right)^2\sigma_n^2 \,.
\label{eq:2}
\end{equation}

A further special case exists if the observables are fully correlated ($\rho= +1$). One obtains

\begin{equation}
\sigma_f^2 = \left( \frac{\partial f}{\partial x_1}\sigma_1 +\frac{\partial f}{\partial x_2}\sigma_2 +\dots + \frac{\partial f}{\partial x_n}\sigma_n\right)^2 \,.
\label{eq:3}
\end{equation}

It is well known that the hadronic contribution to the anomalous magnetic moment of the muon can be calculated from a dispersion integral over the cross sections for the process $e^+ e^- \to \mathrm{hadrons}$~\cite{Bouchiat:1961}. Since KLOE ISR data are binned datasets of $n$ points with binwidth $\Delta s$, the integral can be approximated by a sum. Evaluating $\mathrm{a}_\mu^{\mathrm{had}}$ between $s_{\mathrm{min}}$ and $s_{\mathrm{max}}$, one gets

\begin{equation}
\mathrm{a}_\mu^{\mathrm{had}}[s _{\mathrm{min}},s _{\mathrm{max}}]=\frac{1}{4\pi^3}
\int_{s_{\mathrm{min}}}^{s_{\mathrm{max}}}\sigma^{\mathrm{had}}(s)K(s)\,ds \simeq \frac{1}{4\pi^3} \sum_{i=1}^n \sigma^{\mathrm{had}}_i K_i \Delta s\, .
\label{eq:4}
\end{equation}

Applying Eq.~(\ref{eq:1}) to the sum in Eq.~(\ref{eq:4}) yields

\begin{eqnarray}
\left(\sigma_{\mathrm{a}_\mu^{\mathrm{had}}}\right)^2 &=& \sum_{i=1}^n \sum_{j=1}^n c_i c_j V_{ij} = \underbrace{\sum_{i=1}^n c_i^2 \sigma_i^2}_{\substack{i=j,\\ \mathrm{diagonal\; term}}} + \underbrace{\sum_{i=1}^n \sum_{\substack{j=1\\ j\neq i}}^n c_i c_j \sigma_{ij}}_{\substack{i\neq j,\\ \mathrm{off-diagonal\; term}}} \, ,
\label{eq:5}
\end{eqnarray}

where the $V_{ij}$ are the elements of the covariance matrix and the partial derivatives have been abbreviated as $c_i \equiv {\partial \mathrm{a}_\mu^{\mathrm{had}}}/{\partial \sigma_i}= (1/4\pi^3) K_i\Delta s$. It is immediately obvious that neglecting the term from the off-diagonal elements results in an incorrect value for $\sigma_{\mathrm{a}_\mu^{\mathrm{had}}}$.

Of course, uncertainties that are independent between different data points enter only in diagonal elements of covariance matrix (e.g. the stat. errors of observed spectrum). Recalling Eq.~(\ref{eq:2}), one finds for this class of uncertainties a ``sum of squares'' in the uncertainty estimation of $a_\mu^{\mathrm{\pi\pi}}$:

\begin{equation}
\left(\sigma_{\mathrm{a}_\mu^{\mathrm{\pi\pi}}}\right)^2 =  \sum_{i=1}^n \left(\frac{1}{4\pi^3} (\sigma_{\sigma_{\mathrm{\pi\pi}}})_i K_i \Delta s\right)^2 \, .
\label{eq:6}
\end{equation}

The $(\sigma_{\sigma_{\mathrm{\pi\pi}}})_i$ can be a combination of several uncorrelated uncertainties, like

\begin{equation}
(\sigma_{\sigma_{\mathrm{\pi\pi}}})_i = \sqrt{\left(\sigma_i^\mathrm{spectrum}\right)^2 +\left(\sigma_i^\mathrm{eff}\right)^2 + \dots}
\label{eq:7}
\end{equation}

The $(\sigma_{\sigma_{\mathrm{\pi\pi}}})_i$ are given as the statistical error (properly rounded) attached to the $\sigma_{\pi\pi}$ values in the data tables of the KLOE ISR publications.

Fully correlated uncertainties affect all data points in a similar way, the covariance matrix elements are constructed as $V_{ij} = +1\cdot\sigma_i\sigma_j$ (e.g. the error on int. luminosity measurement $\int{\mathcal{L}}dt$). Using Eq.~(\ref{eq:3}), this time one gets a ``square of sum'' in the uncertainty estimation of $a_\mu^{\mathrm{\pi\pi}}$:

\begin{equation}
\left(\sigma_{\mathrm{a}_\mu^{\mathrm{\pi\pi}}}\right)^2 = \left( \sum_{i=1}^n \frac{1}{4\pi^3} (\sigma_{\sigma_{\mathrm{\pi\pi}}})_i K_i \Delta s\right)^2 \, .
\label{eq:8}
\end{equation}

Unless the uncertainty is constant for all data points, $(\sigma_{\sigma_{\mathrm{\pi\pi}}})_i$ can not be a combined uncertainty. $(\sigma_{\mathrm{a}_\mu^{\mathrm{\pi\pi}}})^2 $ must be evaluated for each individual uncertainty source (background, efficiencies, etc.) and then combined in quadrature (the different sources of uncertainties are independent from each other):

\begin{equation}
\left(\sigma_{\mathrm{a}_\mu^{\mathrm{\pi\pi}}}\right) = \sqrt{\left(\sigma_{\mathrm{a}_\mu^{\mathrm{background}}}\right)^2 + \left(\sigma_{\mathrm{a}_\mu^{\mathrm{eff_1}}}\right)^2 + \left(\sigma_{\mathrm{a}_\mu^{\mathrm{eff_2}}}\right)^2+\dots} \, .
\label{eq:9}
\end{equation}

The fully correlated uncertainties are given as systematic errors in separate tables in the KLOE publications. The dependence of the systematic errors on $s$ is provided in the additional documentation to each analysis, which can be found at~\cite{KLOEurl}.

Partially correlated uncertainties arise in the analyses due to corrections for detector resuolution (``unfolding'') and final state radiation (``unshifting''). Both effects are described in detail in the additional documentation (see e.g.~\cite{KN225}), and result in a redistribution of events between neighboring bins. While for the unfolding, the corresponding entries in the covariance matrix are obtained from the software package used for the Bayesian unfolding approach~\cite{dagostini}, the entries due to the unshifting were derived using the multinomial structure of the underlying transfer matrix. Here we have to rely on Eq.~(\ref{eq:5}) to calculate the effect on $\left(\sigma_{\mathrm{a}_\mu^{\mathrm{\pi\pi}}}\right)$.

Putting everything together and applying Eq.~(\ref{eq:5}) and Eq.~(\ref{eq:9}) (based on Eq.~(\ref{eq:8})) to the data from the KLOE08 analysis~\cite{Ambrosino:2010bv2}, one obtains for $a_\mu^{\pi\pi}$ in the range between 0.1 and 0.85 GeV${}^2$:

\begin{equation}
a_\mu^{\pi\pi}[0.1-0.85 \mathrm{GeV}^2]=(478.5\pm2.0_\mathrm{stat}\pm5.0_\mathrm{exp}\pm4.5_\mathrm{theo})\times 10^{-10}\, ,
\label{eq:10}
\end{equation}

where the fully correlated error has been divided into systematic effects coming from experimental and theoretical sources, and the statistical uncertainty is evaluated plugging the full covariance matrix into Eq.~(\ref{eq:5}). Note that if one had used only the diagonal terms in Eq.~(\ref{eq:5}) (in other words, using only Eq.~{\ref{eq:6})), the statistical uncertainty in Eq.~(\ref{eq:10}) would come out to be lower by 10\%.

The author wishes to thank the organizers of the workshop for the nice atmosphere and ECT* for support.

\newpage

\subsection{Charged $\zc$ at BESIII}
\addtocontents{toc}{\hspace{2cm}{\sl R.G.~Ping}\par}

\vspace{5mm}

R.G.~Ping (for BESIII Collaboration)

\vspace{5mm}

\noindent
Institute of High Energy Physics, Beijing 100049, People's Republic of China\\

\vspace{5mm}

Using a 525 pb$^{-1}$ data sample collected at $\sqrt s=4.26 $ GeV with the BESIII detector, a new charmonium like state, denoted as $\zc$, is observed in the decay $\ee\to\zc^{\pm}\pi^{\mp}\to\ppj$. If the $\zc$ is parametrized as an S-wave Breit-Wigner function convolved with a Gaussian to account for the detector mass resolution, the mass (M=$3889.0\pm3.6\pm4.9$ MeV) and width ($\Gamma=46\pm10\pm20$ MeV) are measured with an unbinned maximum likelihood fit to the distribution of $M(\jp\pi)$ in the maximum mass region event by event \cite{beszc}. The observation of this state was confirmed by the Bell \cite{bellzc} and CLEO-c \cite{cleoczc} experiments.

The Born cross section for $\ee\to\ppj$ is measured to be $62.9\pm1.9\pm3.7$ pb, which is consistent with the existing results from the Babar \cite{babarxs}, Bell \cite{bellxs}, and CLEO \cite{cleoxs} experiments. The production ratio of the $\zc$ is measured to be $R={\sigma(\ee\to\pi^\pm\zc^\mp\to\ppj)\over \sigma(\ee\to\ppj)}=(21.5\pm3.3)\%$.

The $\zc$ observation has triggered some discussions about its structure and properties, such as the tetraquark scheme \cite{tetraquak}, hadronic molecular state \cite{molecular}, meson loop mechanisms \cite{loop}, the initial pion emission mechanism \cite{ipem} and so on. More precise measurements on the $\zc$ mass and width, the quantum number, and production cross sections are desirable and helpful to clarify these schemes.

\newpage

\subsection{ $\gamma \gamma$ Physics at BES-III}
	\addtocontents{toc}{\hspace{2cm}{\sl C.F.~Redmer}\par}

\vspace{5mm}

C.F.~Redmer

\vspace{5mm}
Institut f\"ur Kernphysik, Johannes Gutenberg-Universt\"at Mainz, Germany\\
\vspace{5mm}

The $\gamma \gamma$ physics program~\cite{BES3P} at the BES-III experiment aims at the measurement of the 
electromagnetic transition form factors (TFF) of pseudoscalar mesons in the space-like region. Precise knowledge of the 
TFF is of vital importance for the calculation of the hadronic contribution to the muon anomaly $a_\mu$, which 
completely limits the theoretical prediction of $a_\mu$. The TFF are needed as experimental input in the calculation of
the hadronic light-by-light scattering, since perturbative methods cannot be used in the low energy regime relevant for 
$a_\mu$.

The BES~III experiment~\cite{BES3H} is located at the BEPC~II $e^+ e^-$ collider, operated at the IHEP in Beijing 
(China). Data can be collected in a center-of-mass energy range from 2.0~GeV to 4.6~GeV. For the determination of the 
TFF, data taken at the $\psi(3770)$ peak are being analyzed. Currently, $2.9\textrm{~fb}^{-1}$ have been collected at 
this energy and it is planned to extend the data set to an integrated luminosity of $10\textrm{~fb}^{-1}$.

Feasibility studies~\cite{feas}, performed with the Ekhara~\cite{ekhara,2octet} event generator, show that the TFF for 
$\pi^0, \eta$ and $\eta^\prime$ mesons can be extracted at momentum transfers in the range of \mbox{$0.3 \leq Q^2 [GeV] 
\leq 10$}. Assuming the full integrated luminosity of $10\textrm{~fb}^{-1}$, the statistical accuracy of a TFF 
measurement at BES-III was found to be unprecedented for momentum transfers of $Q^2 \leq 4$~GeV$^2$, a region of special 
relevance for the hadronic light-by-light scattering~\cite{theo1,theo2}. At higher momentum transfers the precision is 
compatible with the CLEO~\cite{CLEO} result, allowing for cross checks with previous measurements of 
TFF's~\cite{OTHER}.

Data analysis is based on a single-tag technique, where only the produced meson and one of the two scattered leptons 
are reconstructed from detector information. The second lepton is reconstructed from four-momentum conservation and 
required to have a small scattering angle, so that the momentum transfer is small and the exchanged photon is 
quasi-real. The ongoing analyses tag the produced pseudoscalar meson in the decay channels 
$\pi^0\rightarrow\gamma\gamma$, $\eta\rightarrow\gamma\gamma$, $\eta\rightarrow\pi^+\pi^-\pi^0$, and 
$\eta^\prime\rightarrow\pi^+\pi^-\eta$. Major sources of background are QED processes such as virtual Compton 
scattering, misidentified hadronic final states, external photon conversion, and on-peak background from two-photon 
processes such as the production of different mesons or and initial state radiation in the signal channel. Conditions 
are being devised to suppress the identified background sources.

\newpage

\subsection{Are isospin correction in $\tau^-\to\pi^-\pi^0\nu_\tau$ decays understood?}
\addtocontents{toc}{\hspace{2cm}{\sl P.~Roig}\par}

\vspace{5mm}

\noindent
P.~Roig

\vspace{5mm}

\noindent
Grup de F\'{\i}sica Te\`orica, Institut de F\'{\i}sica d'Altes Energies,
Universitat Aut\`onoma de Barcelona, E-08193 Bellaterra, Barcelona, Spain.\\

\vspace{5mm}
\hspace{7mm}$\tau^-\to\pi^-\pi^0\nu_\tau$ decays provide an ideal scenario to study the properties of the $\rho^-(770)$ resonance and the hadronization of the vector current 
in presence of strong interactions in the chiral- and intermediate-energy regimes. Moreover, it allows to obtain an alternative evaluation of the $\pi\pi$ contribution 
to the anomalous magnetic moment of the muon ($a_\mu^{\pi\pi}$) \cite{Alemany:1997tn} and (together with the three-prong pion decay mode, see O. Shekhovtsova's contribution) 
is useful to verify the spin and parity of the Higgs boson \cite{Desch:2003rw} discovered at the LHC \cite{Aad:2012tfa}. Belle's data on the di-pion tau decay mode 
\cite{Fujikawa:2008ma} are so precise that they allow to assess whether isospin corrections are understood or not.

In the considered decays, the scalar form factor contribution vanishes even including the leading isospin breaking corrections \cite{Cirigliano:2001er}. Thus, to an excellent 
approximation the involved dynamics can be described just in terms of the vector form factor (VFF), which has been studied extensively in the literature. In particular, its 
expansion at low energies is known up to next-to-next-to-leading order \cite{Gasser:1990bv}. To increase the range of applicability of the previous results the inclusion of 
resonances as explicit degrees of freedom is needed. This is most conveniently done in the antisymmetric tensor formalism, within Resonance Chiral Theory ($R\chi T$) 
\cite{Ecker:1988te}, which has been successfully applied also to the study of other hadronic tau decay modes \cite{HadTauDec, Boito:2008fq}. In Ref.~\cite{Guerrero:1997ku} final 
state interactions were resummed in the resulting VFF by means of an Omn\`es exponentiation. However, analyticity and unitarity constraints were only partially implemented in 
the devised scheme (see also \cite{Pich:2001pj}). A solution to this problem was proposed in Ref.~\cite{Boito:2008fq} for the $K\pi$ VFF. In Ref.~\cite{Dumm:2013zh} we apply 
this construct to the $\pi\pi$ VFF and match it to a phenomenological formula including the contribution of excited resonances, improving \cite{Roig:2011iv} (both from the 
theoretical point of view and from the agreement with data with slightly improves the Gounaris-Sakurai fit made by Belle \cite{Fujikawa:2008ma}), which was in the TAUOLA 2012 
release \cite{Shekhovtsova:2012rap}. The present results will be included in a future upgrade of the generator.

Isospin corrections are implemented accounting for resonance exchange \cite{Cirigliano:2002pv, FloresBaez:2006gf}. These correspond to: the dominant short-distance electroweak 
corrections, the different masses within an isospin multiplet both in the kinematics and in the loop functions entering $\Gamma_\rho(s)$ \cite{GomezDumm:2000fz}, and virtual 
and real photon corrections (distorting the whole spectrum and the form factor, including a local correction with the contribution of the chiral LECs in the latter case). 
Our results show that the inclusion of $SU(2)$ breaking corrections does not improve the quality of the fits. However, the combined error on the isospin breaking corrections 
is only $\sim10\%$ \cite{Fujikawa:2008ma, Cirigliano:2002pv} of the difference between the SM value \cite{Bennett:2006fi} and the theoretical predictions for $a_\mu$ (see 
related contributions in this WG meeting). Our framework is also able to provide good quality fits of $\sigma(e^+e^-\to\pi^+\pi^-)$ at low energies. This will allow us to 
evaluate $a_\mu^{\pi\pi}$ from both $\tau^-$ and $e^+e^-$ consistently fulfilling unitarity, analyticity and the chiral limit of QCD.

\newpage

\subsection{Analysis of $\eta$ and $\eta'$ Transition Form Factors with rational approximants}
\addtocontents{toc}{\hspace{2cm}{\sl P.~Sanchez-Puertas}\par}

\vspace{5mm}

P.~Sanchez-Puertas

\vspace{5mm}

\noindent
Institut f\"ur Kernphysik, Johannes Gutenberg Universt\"at Mainz, Germany\\

\vspace{5mm}

Pseudoscalar Transition Form Factors (TFF) have received a lot of attention in the QCD community during last years. Theoretically, the limits $Q^2=0$
and $Q^2\rightarrow\infty$ are well known in terms of the axial anomaly in the chiral limit of QCD \cite{Adler:1969gk,Bell:1969ts} and perturbative QCD (pQCD)
\cite{Lepage:1980fj} respectively. On the other hand, despite many efforts have been done to describe the mid-energy behavior, this is not well understood yet.
One of the most common approaches is to extend the pQCD description to these energies, matching the result at $Q^2=0$. Then the problem is reduced
to obtain a description for the pseudoscalar distribution amplitude (DA)~\cite{Mueller:1994cn}. However, it is not certain whether non-perturbative effects are
small enough at these energies to apply pQCD. Furthermore, obtaining the DA is not possible from first principles. Then, some model
needs to be used either for the DA or the TFF itself. The discrepancy among different approaches reflects the model-dependency of the results.\\

Then, we adopt a different approach and try to obtain a good description for the space-like TFF through a fitting procedure to data. For this, we extend~\cite{EscribanoMasjuan} the
use of Pad\'e Approximants (PA) as fitting functions described in Ref.~\cite{Masjuan:2012wy} to the $\eta(')$ system.
As explained in Refs.~\cite{Masjuan:2012wy,Masjuan:2008fv}, this systematic approach allows a model-independent extraction of the low energy (LE) parameters.
Two applications (the estimate of HLBL to $(g-2)_{\mu}$ and $\eta-\eta'$ mixing parameters) are described below.\\

Using our description of the TFF we are able to constrain models accounting for the hadronic light-by-light contribution (HLBL) to the $(g-2)_{\mu}$ along the lines of
Refs.~\cite{deRafael:1993za,Knecht:2001qf,Masjuan:2012wy} in the pseudoscalar pole approximation (see discussions concerning
pseudoscalar off-shellness~\cite{Melnikov:2003xd,Jegerlehner:2009ry}). The novelty in our approach is twofold, on one side we provide a
systematic description of the TFF in the framework of large-$N_c$ theories~\cite{Masjuan:2012qn}, while, in the other, we use our estimate of the LE parameters to constrain
this TFF and provide a systematic error. Our preliminary result is $a^{HLBL;\eta}_{\mu}=1.38(10)_{stat}(7)_{sys}\times10^{-10}$ and
$a^{HLBL;\eta'}_{\mu}=1.26(8)_{stat}(6)_{sys}\times10^{-10}$. Including the previous result for the $\pi^0$~\cite{Masjuan:2012wy}, we estimate for the pseudoscalar contribution

$$ a^{HLBL;PS}_{\mu}=8.2(5)_{stat}(4)_{sys}\times10^{-10}.$$

Finally, the physical $\eta$ and $\eta'$ mesons are an admixture of the SU(3) Lagrangian eignestates~\cite{Leutwyler:1997yr,Feldmann:1998vh,Feldmann:1999uf,Escribano:2005qq}.
Deriving the parameters governing the mixing is a challenging task. Usually, these are determined through the use of $\eta(')\rightarrow2\gamma$ decays as well as vector radiative
decays into $\eta(')$ (see Refs.~\cite{Feldmann:1998vh,Escribano:2005qq}). However, since pQCD predicts
that the asymptotic limit of the TFF for the $\eta(')$ is essentially given in terms of these mixing
parameters~\cite{Lepage:1980fj,Feldmann:1999uf}, we propose to use our TFF parametrization to estimate the asymptotic limit and further constrain the mixing parameters.

\newpage

\subsection{Comparison of Resonance Chiral Lagrangian results with $\tau^-\to \pi^+\pi^-\pi^-\nu_\tau$ BaBar data}
\addtocontents{toc}{\hspace{2cm}{\sl O.~Shekhovtsova}\par}

\vspace{5mm}

I. Nugent$^{a}$, T. Przedzi\'nski$^{b}$,
P. Roig$^{c}$, O. Shekhovtsova$^{d,e}$\footnote{Speaker, IFJPAN-IV-2013-6}, and Z. W\c{a}s$^{e,f}$

\vspace{5mm}

\noindent
 {\em $^a$ III. Physikalisches Institut B RWTH Aachen, Aachen, Germany}\\
{\em $^b$ 
Jagellonian University, Cracow, Poland}\\
{\em $^c$ 
Universitat Aut\`onoma de Barcelona, Barcelona, Spain} \\
{\em $^d$ Kharkov Institute of Physics and Technology, Kharkov, Ukraine}\\
 {\em $^e$  Institute of Nuclear Physics, Krak\'ow,  Poland}\\
{\em $^f$ CERN PH-TH, CH-1211 Geneva 23, Switzerland }

\vspace{5mm}

Investigating $\tau$ lepton processes can present
broad physics interest because of its long lifetime, large mass
and parity sensitive couplings. From the perspective of
high-energy experiments such as those at LHC, good understanding
of tau leptons properties provides important clues of new physics
signatures. With the discovery of a new particle around the mass
of 125-126GeV \cite{higgs},
tau decays are an important
tool for determining if the particle predicted by the Standard Model.
 From the perspective of lower energies,
$\tau$ lepton decays constitute an
excellent laboratory for studies of hadronic interaction at
the energy scale of about 1 GeV, where neither perturbative QCD methods nor chiral Lagrangians are
expected to work to a good precision \cite{Braaten:1990ef}.
At present, hundreds of milions of $\tau$ decays have been amassed by
both Belle and BaBar experiments and most of these data samples are not yet analyzed.

The first version of the program TAUOLA \cite{Jadach:1993hs} was written in the 90's.
TAUOLA is Monte Carlo event generator which simulates tau decays for
both leptonic and hadronic decays modes. The hadronic currents
implemented at that time in TAUOLA are based on Vector Dominance Model \cite{Kuhn:1992nz}.
In this framework the hadronic current for a three-pseudoscalar
decay mode is a sum of weighted products of Breit-Wigner functions.  It is demonstrated in \cite{GomezDumm:2003ku}
that this approach is able to reproduce only leading order $\chi$PT results, and,
moreover, the model is not sufficient to describe the
Cleo $KK\pi$ data \cite{Coan:2004ep}. The alternative approach based on Resonance Chiral Lagrangian (R$\chi$L) 
was proposed in  \cite{GomezDumm:2003ku}. The calculations done within R$\chi$L are able to
reproduce the low-energy limit of $\chi$PT at least up to next-leading order and show the right falloff of hadronic form factors in the high energy region.

In our paper \cite{Shekhovtsova:2012ra} we have described an upgrade of the Monte Carlo generator TAUOLA using the results of R$\chi$L for the tau lepton decay into the most important two and three mesons.
The necessary theoretical concepts were summarized and
numerical tests of the implementations were completed and documented. The manual as well as the differential spectra  and numerical tests are available at \cite{rchl_cite}.
The updated version of the generator TAUOLA
can be incorporated into software enviroments of both BaBar and Belle collaborations as well as of FORTRAN and C++ applications of LHC.

To fit the model parameters we started with the $\pi^-\pi^-\pi^+$ mode.
We would like to stress here that the choice of the three pion mode is not accidental. The kinematical configuration is complex and the three pion mode
has the largest branching ratio among the three meson decay modes. Moreover, this decay mode together with the decays into two pions, which is much easier to model, are a useful tool
for spin-parity analysis of the recently discovered Higgs boson 
\cite{higgs} through its di-tau decays \cite{Czyczula:2012ny}.
The first comparison of the R$\chi$L results for the $\pi^-\pi^-\pi^+$ mode with the BaBar data
\cite{Nugent:2013ij} was done in \cite{Shekhovtsova:2013rb}. Both the three particle and the $\pi^+\pi^-$ invariant
mass distributions were considered. Disagreement at about 12\% level is visible
in the low energy region of two-particle invariant mass distribution whereas for the three-particle invariant mass spectrum the difference between MC and data is less than 7\%.
The fit was done taking into account only $P$-wave contribution of two pion system.
As suggested in \cite{Shibata:2002uv} the discrepancy in the low mass region could be reduced resorting to a contribution
from the scalar particle, $S$-wave contribution.

With the recent availability of the unfolded distributions for all
invariant masses constructed from observable decay products for
this channel \cite{Nugent:2013ij}, we found ourselves in excellent
position to work on model improvement. Previously missing
contribution from the $\sigma$ resonance was added to our currents
and final state Coulomb interactions were taken into account. As a
result, we improved the agreement with the data by a factor of
about three, see \cite{rchla1} in web-page \cite{rchl_cite}.
Remaining differences are well below uncertainties expected for
the  R$\chi$L currents approach. This is the first case when
agreement for a non-trivial $\tau$ decay channel was obtained for
the chiral inspired approach. The R$\chi$L currents are ready for
comparison with data for other $\tau$ decay channels and for work
when unfolded multi-dimensional distributions are used.

\newpage

\subsection{Muon $g-2$ and QCD sum rules}
\addtocontents{toc}{\hspace{2cm}{\sl H.~Spiesberger}\par}

\vspace{5mm}

\noindent
H.~Spiesberger

\vspace{5mm}

\noindent
PRISMA Cluster of Excellence, Institut f\"{u}r Physik,
\\
Johannes Gutenberg-Universit\"{a}t,
Staudingerweg 7, D-55099 Mainz, Germany

\vspace{5mm}
We have analyzed the lowest order hadronic contribution to the
$g-2$ factor of the muon in the framework of the operator product
expansion (OPE) at short distances \cite{Bodenstein:2013flq}.
$e^+ e^-$ data contributing to the leading-order hadronic part
of $a_\mu$ are least precise in the energy range between $s=1$
and $s=(1.8)^2$ GeV$^2$. We have used a quenched finite energy
sum rule to optimize the suppression of data in this energy range.
This procedure reduces the discrepancy between experiment and theory,
$\Delta a_\mu \equiv a^{EXP}_\mu - a^{SM}_\mu$, from
$\Delta a_\mu = 28.7 (8.0) \times 10^{-10}$
\cite{Davier:2010nc,Hagiwara:2011af} to
$\Delta a_\mu = 19.2 (8.0) \times 10^{-10}$.

On the one hand, the error due to uncertainties in the data is
reduced, but additional uncertainties enter the OPE contribution
and are due to (i) the error on the strong coupling constant
$\alpha_s(\mu)$; (ii) an estimate of the error of the perturbative
QCD part, obtained by comparing predictions at ${\cal O}(\alpha_s^4)$
and ${\cal O}(\alpha_s^3)$; (iii) the prescription to perform the
contour integral, i.e.\ fixed-order or contour-improved perturbation
theory; (iv) the value of the condensates, where the gluon condensate
dominates. The combination of these uncertainties turns out to
be the same as in the standard approach which is based exclusively
on data.

We have checked that a presently available model for duality
violations \cite{Cata:2008ru} can not explain the observed
discrepancy between the data-based and OPE-based approaches.
While there is no gauge-invariant dimension-2 condensate, it
has been speculated \cite{Chetyrkin:1998yr} that there might
exist a gauge non-invariant, effective condensate of dimension
2. However, if such a term would be present in the OPE, it would
lead to a further significant decrease of the discrepancy
$\Delta a_\mu$.

We have based our numerical analysis on the set of $e^+e^-$ data
described in \cite{Bodenstein:2011hm}. With this choice, $a_\mu$
is dominated by data from BaBar \cite{Lees:2012cj}. Replacing
the $2\pi$ BaBar data by the somewhat less precise data from KLOE
\cite{Ambrosino:2010bv}, we find a similar shift. We conclude
that the result of our analysis is not an artifact of the data
selection and the observed tension between the OPE and the
$e^+ e^-$ data imply a problem with the OPE and/or with the
$e^+ e^-$ data.

\newpage

\subsection{Radiative corrections to $\bar p+p\to e^+ + e^-$}
\addtocontents{toc}{\hspace{2cm}{\sl E.~Tomasi-Gustafsson}\par}

\vspace{5mm}

\noindent
E.~Tomasi-Gustafsson

\vspace{5mm}

\noindent
CEA,IRFU,SPhN, Saclay, 91191 Gif-sur-Yvette, and \\
Univ Paris-Sud, CNRS/IN2P3, IPNO, UMR 8608, 91405 Orsay, France\\

\vspace{5mm}
This contribution has the purpose to stress the importance of radiative corrections (RC) in other reactions than those studied at $e^+e^-$ colliders, as $ep$ (elastic and inelastic) scattering and particularly in the $\bar p+p\to e^++e^-$ reaction, which is foreseen at PANDA (FAIR) \cite{Panda}.

Although the required degree of precision is one (or more) order of magnitude lower than in experiments for the determination of the hadronic to leptonic cross section $R$, or for calculating the relevant corrections to $(g-2)$, it is important that experimentalists get fully aware that QED RC change not only the magnitude of the observables, but also their dependence on the relevant kinematical variables.

In experiment as DVCS or $ep$ elastic scattering a precise calculation of these corrections appears essential before extracting any information on the proton structure \cite{Bytev:2003qf,Bystritskiy:2006ju}.
The emission from the proton is usually neglected, however its interference with the electron emission may have serious consequences on aspects as angular asymmetries and nonlinearities in the relevant distributions. Concerning form factors (FFs) extraction, attention was driven on final state radiation (FSR) in $e^++e^-\to \bar p+p(\gamma)$ in Ref. ~\cite{Bytev:2011pa}.

An important aspect, which makes this problem very timely is that, differently from the past, present experiments are mostly multi-particle detection and coincidence experiments, with large coverage detectors, both in angle and in momentum acceptance. Therefore a (precise) calculation of RC should be complemented by a Monte Carlo application, embedded in the analysis programs. This arises numerical and formal problems of divergences and instabilities. It appears to me that a structure as this MonteCarLow group, gathering experts, experimentalists and theoreticians is the right way to proceed and perform these analysis.

Concerning $\bar p+p\to e^+ + e^-$, an analysis with PHOTOS package \cite{Golonka:2005pn}  as implemented in PANDARoot (the software of the PANDA experiment both for simulation and analysis) was performed and compared to the existing calculations \cite{Ahmadov:2010ak,VandeWiele:2012nb}. PHOTOS allows to apply the correction to simulated events from final state radiation (FSR), in Leading Log approximation (LLA). The feature that these RC depend on the electron angle in Lab system \cite{Gakh:2011mn} is reproduced by PHOTOS.

The initial state radiation (from proton-antiproton), as well as two photon exchange (which contribute at the level of $\alpha^3$) are not taken into account in PHOTOS. The vacuum polarization is also not included, but can be considered as global normalization. The radiation from the (anti)proton is in principle small with respect to electron emission, as it inversely proportional to the mass of particle. However, for the extraction of FFs, it is very important because it induces an odd-$\cos\theta$ contribution, whereas the formalism for extraction of the FFs moduli is based on one photon-exchange formalism, Ref.~\cite{Zichichi:1962ni}, which predicts that the cross section is an even function of $\cos\theta$. Any angular asymmetry, even at the level of few percent, if not taken into account, will make the extraction of the FFs false. An analysis based on symmetry properties of the strong and electromagnetic interaction suggests how to single out or to cancel such asymmetries \cite{Gakh:2005wa} (\cite{Gakh:2005hh} for the time-reverse process).

The values of the different contribution to RC from the available calculations \cite{Ahmadov:2010ak,VandeWiele:2012nb} and as extracted from the PHOTOS-PANDARoot simulation were compared for $\bar p+p\to e^+ + e^-$ for $s=12.9$ GeV$^2$ and with a cut on the photon energy, $E_\gamma<100$ MeV.

RC strongly depend on the experimental cuts, and particularly on the hard photon cut. A realistic value of this cut will be finally settled when the detector capability to detect the radiated photons with a specific resolution will be demonstrated on real data.

The results given by the simulation are of the order of $18\%$ and are consistent with the analytical calculations for the even contributions, excepted the vacuum polarization. Odd contributions arising from the interference between the initial and the final state
(and, at a lower extent, from the two photon contribution) are not calculated by PHOTOS. Such contributions are quite large (8\%) and result in a distortion of the angular distribution affecting the extraction of form factors and increasing the systematic errors. Simulations \cite{Sudol:2009vc} show that PANDA will detect a $\cos\theta$ asymmetry at the level of $5\%$ or more. The question to be elucidated, possibly in frame of these series of meeting, is if it is more convenient to include the missing terms in PHOTOS or to prepare a specific generator for the FFs extraction in $\bar p+p\to e^+ + e^-$.

\newpage

\subsection{Looking for the phase angle between strong and EM mechanisms in $\jpsi$ decays}
\addtocontents{toc}{\hspace{2cm}{\sl Y.~Wang}\par}

\vspace{5mm}

\noindent
Yadi Wang \\ on behalf of the BES~III (LNF-INFN) group

\vspace{5mm}

\noindent
Laboratori Nazionali di Frascati (INFN) (Rome), Italy\\

\vspace{5mm}
The $\jpsi$ meson has been found more than 40 years ago. It can decay into hadrons mainly via strong or EM process. The 93 KeV decay width is often explained as a proof of the application of pQCD. In pQCD, the $\jpsi$ strong amplitude $A_{3g}$ and the EM amplitude $A_{EM}$ are predicted to be both real~\cite{Chernyak:1984bm,Bolz:1996sw,Brodsky:1987bb}, as expected for the non-resonant amplitude $A_{cont.}$. In view of QCD, the phase angle between $A_{3g}$ and $A_{EM}$ is no more than $10^{\circ}$~\cite{Brodsky:1987bb}. The interference between $\jpsi$ EM decay amplitude and non-resonant decay amplitude has been observed in $\elp\elm\to\mup\mum$ at SLAC~\cite{Boyarski:1975ci}, BESII~\cite{Bai:1995ik}, KDER~\cite{Anashin:2009pc}, which is in good aggreement with what expected. On the other side, a series of experimental evidence ($\jpsi\to N\bar{N}, VP, PP, VV$, where $N$, $V$ and $P$ stand for nucleon, vector and pseudo-scalar meson repectively)~\cite{Baldini:1996hc,Baldini:1998en,Kopke:1988cs,Jousset:1988ni,Suzuki:1999nb,Ablikim:2012eu} point toward an unexpected $\sim 90^{\circ}$ phase difference. That means the $\jpsi$ strong decay amplitude is orthogonal to the $\jpsi$ EM decay amplitude. All the imaginary amplitudes are mostly obtained by comparing decay processes, belonging to the same category, modelling the amplitudes by means of $SU_{3}$ and $SU_{3}$ breaking. These additional theoretical hypotheses are questionable~\cite{Chernyak:1999cj}. 

Another model-independent way to measure the phase difference between the strong and EM amplitudes is by searching for an interference in the $Q^{2}$ behavior, since the interference term should vanish once inclusive processes are summed up. The overall cancellation could be achieved by oppsite real amplitudes. Thus, the investigation of exclusive channels is nessary. 

From another point of view, not all the interferences between strong and EM mechanisms behave the same way in quarkonium decays. For example, unlike $c\bar{c}$, in the $s\bar{s}$ bound state $\phi$ decays to $\rho\pi$ process, an $180^{\circ}$ phase difference was found in Novosibirsk experiment. Thus, it would be interesting to see the interference beheavior of strong and EM mechanisms in $\jpsi$ decays.

BES~III/BEPC~II~\cite{2009vd} is a major upgrade of the BESII detector and BEPC accelerator. The primary physics purposes are aimed at the study of hadron spectroscopy and $\tau$-charm physics.
A fine scan around $\jpsi$ peak has been finished in 2012. Based on this data sample, the analysis of $\elp\elm\to\mup\mum$, $\elp\elm\to 2(\pip\pim)$ and $\elp\elm\to 2(\pip\pim)\piz$ and so on for the phase difference measurement are under going and will, hopefully, allow to better understand the $\jpsi$ decay mechanism.

\newpage

\newpage

\section{List of participants}

\begin{flushleft}
\begin{itemize}
\item H.~Czy\.z, University of Silesia, {\tt henryk.czyz@us.edu.pl }
\item M.~De Stefanis, Universit\`a degli Studi di Torino, {\tt destefan@to.infn.it}
\item A.~Denig, Universit\"at Mainz, {\tt denig@kph.uni-mainz.de}
\item S.~Eidelman, Novosibirsk State University, {\tt eidelman@mail.cern.ch }
\item K.~Griessinger, Standford University, {\tt griess@slac.stanford.edu}
\item M.~Gunia, University of Silesia, {\tt guniamichal@gmail.com}
\item A.~M.~Hafner, Universit\"at Mainz, {\tt hafner@slac.stanford.edu }
\item H.~Hu, Institute of High Energy Physics, Beijing, {\tt  huhm@ihep.ac.cn }
\item F.~Jegerlehner, Humboldt -Universit\"at, Berlin and DESY, Zeuthen, {\tt fjeger@physik.hu-berlin.de }
\item T.~Johansson, Uppsala University, {\tt tord.johansson@physics.uu.se}
\item B.~Kloss, Universit\"at Mainz, {\tt kloss@uni-mainz.de }
\item K.~Kolodziej, University of Silesia, {\tt karol.kolodziej@us.edu.pl}
\item J.~K\"uhn, Institut f\"ur Theoretische Teilchenphysik, {\tt Johann.Kuehn@KIT.edu}
\item A.~Kupsc, Uppsala University, {\tt Andrzej.Kupsc@physics.uu.se }
\item E.~A.~Kuraev, Bogoliubov Laboratory of Theoretical Physics, {\tt kuraev@theor.jinr.ru}
\item P.~Masjuan, Universt\"at Mainz, {\tt masjuan@kph.uni-mainz.de}
\item S.~.E.~M\"uller, Helmholtz-Zentrum Dresden-Rossendorf, {\tt stefan.mueller@hzdr.de}
\item V.~Pauk, Universit\"at Mainz, {\tt paukvp@gmail.com }
\item A.~Penin, University of Alberta, {\tt penin@ualberta.ca}
\item C.~Redmer, Universit\"at Mainz, {\tt redmer@uni-mainz.de }
\item M.~Ripka, Universit\"at Mainz, {\tt ripka@uni-mainz.de}
\item P.~Roig, Universitat Aut\`onoma de Barcelona, {\tt proig@ifae.es }
\item P.~RongGang, Institute of High Energy Physics Beijing, {\tt pingrg@ihep.ac.cn }
\item P.~Sanchez-Puertas, Universit\"at Mainz, {\tt sanchezp@kph.uni-mainz.de }
\item O.~Shekhovtsova, Institute of Nuclear Physics Cracow, {\tt Olga.Shekhovtsova@lnf.infn.it }
\item H.~Spiesberger, Universit\"at Mainz, {\tt spiesber@uni-mainz.de }
\item E.~Tomasi-Gustafsson, CEA,IRFU,SPhN, Saclay, {\tt etomasi@cea.fr}
\item T.~Teubner, University of Liverpool, {\tt thomas.teubner@liverpool.ac.uk }
\item M.~Unverzagt, Universit\"at Mainz, {\tt unvemarc@kph.uni-mainz.de }
\item M.~Vanderhaeghen, Universit\"at Mainz, {\tt marcvdh@kph.uni-mainz.de }
\item G.~Venanzoni, Laboratori Nazionali di Frascati dell'INFN, {\tt Graziano.Venanzoni@lnf.infn.it }
\item Y.~Wang, Laboratori Nazionali di Frascati dell'INFN, {\tt wangyd@lnf.infn.it }
\item B.~Zhang, Institute of High Energy Physics, Beijing, {\tt zhangbx@ihep.ac.cn}
\end{itemize}
\end{flushleft}

\end{document}